\documentstyle[aps,twocolumn,graphicx,psfrag]{revtex}
\begin{document}
\title{Rotating Spin-1 Bose Clusters}
\author{Tin-Lun Ho and Erich J Mueller}
\address{Department of Physics,  The Ohio State University,
Columbus, Ohio
43210}
\date{Revised March 17,2002 Mueller}

\wideabs{
\maketitle

\begin{abstract}
We propose a simple scheme for generating rotating atomic
clusters in an optical lattice which produces states
with quantum Hall and spin liquid properties.
As the rotation frequencies increases,
the ground state of a rotating cluster of spin-1 Bose atoms
undergoes a sequence of (spin and orbit) transitions,
which terminates at an angular momentum $L^{\ast}$ substantially lower than
that of the boson Laughlin state.
The spin-orbit correlations reflect `fermionization' of bosons
facilitated by their spin degrees of freedom. We also show that
the density of an expanding group of clusters has a scaling form
which reveals the quantum Hall and the spin structure of
a single cluster.
\end{abstract}
}

In this paper, we would like to point out some remarkable features of
rotating clusters of spin-1 Bose atoms which result from the coupling
of quantum Hall physics to spin symmetry through Bose statistics.
Our studies are prompted by the recent experiments on rotating Bose
gases\cite{Paris} and Bose Hubbard systems\cite{Germany}, as well as the
theoretical realization of a `fragmented' condensate in
spin-1 Bose gas\cite{HoYip,Law}, and the theoretical demonstration that a
three  dimensional Bose system can approach the quantum Hall limit at
sufficiently large angular momenta\cite{Horot}. As of now, one can release the
spin degrees of freedom of a Bose gas in an optical trap\cite{MIToptical},
 deposit a large amount of angular momentum to a Bose
condensate\cite{Paris}, and divide a Bose condensate into thousands of
isolated clusters each with a few particles\cite{Germany}. It is natural to
ask whether these capabilities can be combined to produce a large
collection of rotating and spin carrying clusters. In fact, such possibilities
are being considered in some laboratories\cite{Bloch}.

There are many reasons for studying these Bose clusters: Firstly,
they are the systems where two types of strongly correlated ground
states can be realized: quantum Hall droplets and fragmented Bose
spin liquids. While the former is familiar to condensed matter
physicists, the latter is new, with properties entirely different
from the familiar single condensate (or coherent) states\cite{HoYip}.
It is well
known, however, that  these states are very fragile as number of
particle increases, easily giving way to the highly robust coherent
states\cite{HoYip,comment1}. On the other hand, such fragility  disappears in
small clusters, and the system becomes strongly correlated in this
regime.  Secondly,  quantum clusters are amenable to exact treatments,
which may reveal physical principles operative in larger scale.
Thirdly, amid the growing interest in using bosons in lattices to
process quantum information, the study of quantum clusters remains a
basic step.

The results we will present are: a simple method to generate
rotating clusters, the general features of their
ground states, and the scaling behavior of the density of an expanding
group of clusters, whose specific form reflects
spin liquid and quantum Hall properties.
Before proceeding, we first summarize the general features:

\noindent {\bf (A):} As the angular momentum $L$ of the system
increases, the interaction energy will be `quenched' to zero at a
critical value $L^{\ast}$ considerably lower than that of the boson
Laughlin state.
\noindent {\bf (B):} As $L$ increases
from 0 to $L^{\ast}$, the total spin of the ground state is
specified by its orbital angular momentum,
even though the interaction is invariant under separate
spin and orbital rotation.
 {\bf (C):} Both {\bf (A)} and {\bf (B)} reflect the phenomenon
of `fermionization' facilitated by the internal degrees of the freedom
(i.e. spin) of the system.  This is achieved  by
populating bosons in different {\em spin-orbit} states, a process
automatic for fermions.
\noindent  {\bf (D):} The energetics leading to {\bf (A)} and {\bf (B)}
also  implies that as the rotational frequency $\Omega$ increases toward
the  trap frequency $\omega$, the angular momentum of the ground
state will  increase in discrete steps until it reaches $L^{\ast}$, then
remains constant until the system becomes unstable
at $\Omega>\omega$.
 We shall now derive these results.

{\bf I. Rotating quantum clusters: their generation and Hamiltonian:}
Consider adding a rotating quadrupolar potential $W({\bf r};t) = 
\lambda(r_{1}^2(t) - r_{2}^2(t))$, where $r_{i}(t)={\bf r}\cdot \hat{\bf
e}_{i}(t)$, $\hat{\bf e}_{1}(t) = \hat{\bf x} {\rm cos}\Omega t 
+ \hat{\bf y} {\rm sin}\Omega t $,  $\hat{\bf e}_{2}(t) = \hat{\bf y} {\rm
cos}\Omega t  - \hat{\bf x} {\rm sin}\Omega t $, and $\lambda$ is a constant. 
Such a potential can be generated by a pair of off-center
rotating laser beam as in the Paris experiment\cite{Paris}.
The single particle Hamiltonian is then
$h(t)  = \hat{T} + \hat{U} +W({\bf r};t)$, where $\hat{T}={\bf p}^2/2M$, and 
$\hat{U}$ is the periodic potential of the optical lattice. Deep in the Mott
limit, $\hat{U}$ is  an array of deep wells at lattice sites
$\{ {\bf R} \}$; the bottom of each is harmonic,
$\hat{U}({\bf r}) \rightarrow \hat{U}_{\rm har}({\bf x}) = 
\frac{1}{2}M\omega^{2}{\bf x}^2$ as ${\bf r}\rightarrow {\bf R}$, 
${\bf r} = {\bf x} + {\bf R}$.
These wells are so deep that  atoms in different wells are
isolated from each other \cite{stable}. 
Shifting the origin to ${\bf R}$, the
single particle Schrodinger equation is 
$i\hbar\partial_{t}|\Psi_{t}\rangle =
[\hat{T} + \hat{U}_{\rm har}({\bf x}) + 
W({\bf x} + {\bf R};t)] |\Psi_{t}\rangle$. 
Defining 
$|\Psi_{t}\rangle$=
$ e^{-i[{\bf a}(t)\cdot{\bf p} + {\bf b}(t)\cdot {\bf x}]/\hbar} $
$ e^{-i\theta (t)}$ $e^{-i\Omega t \hat{L}_{z} }$
$|\Phi_{t}\rangle$, and with appropriate choices of ${\bf a}$,
${\bf b}$ and $\theta$ \cite{choices}, 
the Schrodinger equation can be written 
as $i\hbar\partial_{t}|\Phi_{t}\rangle = \overline{\cal K} |\Phi_{t}\rangle$, 
where
$\overline{\cal K} = \hat{T} + \hat{U}_{\rm har}({\bf x}) +
W({\bf x};0) -\Omega L_{z}$ is time independent, and $L_{z}= xp_{y}- yp_{x}$.
Since $W({\bf x},0)$ breaks cylindrical symmetry, it will impart angular 
momentum to the system. 
Once the system acquires angular momentum,  
cylindrical symmetry can be restored by turning off $W$, and 
$\overline{\cal K}$ reduces to the simple form 
${\cal K} ={\bf p}^2/2M + \frac{1}{2}M\omega^2 {\bf x}^2 -
\Omega L_{z}$.

Including particle interactions, the Hamiltonian of a rotating spin-1
Bose cluster is, in first quantized form,  $\hat{H}-\Omega \hat{L}_{z} =
\hat{\cal K} + \hat{\cal V}$, where $\hat{\cal K}= \sum_{i}
\hat{\cal K}_{i}$,
$\hat{\cal V} = \sum_{i>j} \hat{\cal V}_{ij}$,
$\hat{\cal V}_{ij}$ $ = $ $( c_{o} +  c_{2} {\bf F}_{i}$ $\cdot$
${\bf F}_{j})$ $ \delta$ $ ({\bf r}_{i}- {\bf r}_{j})$,
where $c_{o}$ and $c_{2}$ are density-density
and spin-spin interactions\cite{Hospinor,Japan}.
Theoretical estimates show that
$c_{o}>0$  for both $^{23}$Na and $^{87}$Rb; and that
$c_{2}>0$ (antiferromagnetic) for  $^{23}$Na and $c_{2}<0$
(ferromagnetic) for $^{87}$Rb\cite{Hospinor}. Because of the similarity
of scattering lengths in different angular momentum channels, it is
found that $|c_{2}|/c_{o}\sim 5\%$ for
both cases.  From now on, we shall focus on values of $c_0$ and $c_2$
appropriate
for $^{23}$Na and $^{87}$Rb.
For simplicity, we
first consider zero magnetic field. Magnetic field effects
will be considered at the end.

The eigenvalues of $\hat{\cal K}$ are\cite{Horot}
$E_{n,m ; n_{z}}/\hbar=  (\omega + \Omega)n + (\omega - \Omega)m +
\omega n_{z}$, where  $n$, $m$, and $n_{z}$ are non-negative integers.
For very tight traps,  $\hbar \omega> Nc_{o}/a^3$ 
with $a=\sqrt{\hbar/M\omega}$,
only  states in the `lowest Landau level' (LLL), $(n=0,m;n_z=0)$,
will appear in  the ground states.
The wavefunction of the
state $(0,m; 0)$ is $w_{m}(x,y)f_{0}(z)$, where
$w_{m}(x,y) = u^{m} e^{-|u|^2/2}/\sqrt{\pi a^2 m!}$,
$f_{0}(z)= e^{-z^2/2a^2}/(\pi
a^2)^{1/4}$, and  $u= (x+iy)/a$.  The many body
state is then
\begin{equation}
|\Phi\rangle = \overline{\int}\!\! B\left([u], [\mu]\right)\,\prod_{i=1}^{N}
\hat{\psi}^{\dagger}_{\mu_{i}}({\bf r}_{i})\,|0\rangle
\label{Phi} \end{equation}
where $\overline{\int} (^{...})  \equiv \int \prod_{i=1}^{N} [{\rm d}{\bf
r}_{i}\,
 f_{0}(z_{i})e^{-|u_{i}|^2/2}] (^{...})$, $[u]\equiv (u_{1}, u_{2}, ...
u_{N})$, $\mu_{i}=1,0, -1$ labels the spin of the $i$-th boson,
$[\mu]\equiv
(\mu_{1},  \mu_{2}, ... \mu_{N})$,
$B([u],[\mu])$ is a symmetric homogeneous polynomial of $[u]$ labelled
by
$[\mu]$,   $\hat{\psi}({\bf r})_{\mu}$ is the field operator in the
LLL,
$\hat{\psi}_{\mu}^{\dagger}({\bf r})= \sum_{m} w_{m}(x,y)f_{0}(z)
a^{\dagger}_{m\mu}$, and
$a^{\dagger}_{m\mu}$ is the creation operator for the state $(0,m;0)$.
In LLL, we have $\hat{\cal K}=\hbar (\omega - \Omega)
\hat{L}_{z}$,
$\hat{L}_{z} = \sum_{m=0}^{\infty} m a^{\dagger}_{m \mu} a^{}_{m\mu}$.
Since $[\hat{\cal V},\hat{L}_{z}]=0$, the ground state of
the Hamiltonian $\hat{H}-\Omega \hat{L}_{z}$
can be determined once the spectrum of $\hat{\cal V}$ as a
function of $L_{z}$ (${\cal V}_{L}$) is obtained.

Before proceeding, we define
$A^{x}({\bf r})= [-\hat{\psi}_{1}({\bf r})+
\hat{\psi}_{-1}({\bf r})]/\sqrt{2}$,
$A^{y}({\bf r}) = -i[\hat{\psi}_{1} ({\bf r}) +
\hat{\psi}_{-1} ({\bf r})]/\sqrt{2}$,
$A^{z}({\bf r}) = \hat{\psi}_{0} ({\bf r})$. Under a spin rotation,
$\vec{A}$ rotations as a 3D vector in $(xyz)$ space.
Thus, states with total spin ($S=0$, 1, and 2)  formed by two
bosons at ${\bf r}_1$ and ${\bf r}_2$ are :
$\vec{A}^{\dagger}_{1}\cdot \vec{A}^{\dagger}_2$,
$\vec{A}^{\dagger}_{1}\times \vec{A}^{\dagger}_2$,  and
$A_{i1}^{\dagger} A^{\dagger}_{j2} -
\frac{1}{3}\delta_{ij}
\vec{A}^{\dagger}_{1}\cdot \vec{A}^{\dagger}_{2}$, where
$\vec{A}_{1}\equiv \vec{A}({\bf r}_{1})$.

{\bf II. The role of internal degrees of freedom and the
`quenching' angular momentum $L^{\ast}$:}
We have numerical diagonalized  $\hat{\cal V}$  up to $N=5$ particles.
The spectrum ${\cal V}_{L}$ of the $N=4$ cluster with $c_{2}>0$ is shown
in figure 1. It shows  a clear correlation between spin and orbital angular
momentum in the ground state. In the following, we shall explain the
origin, the systematics, and the analytic structures of these states.

Since $c_{o}>|c_{2}|$, the  minimum
of the interaction $\hat{\cal V}_{ij}$ is zero, which
occurs if the  relative angular momentum $\ell$ between bosons $i$ and
$j$ is non-zero.
 For scalar bosons, Bose statistics demands $\ell \geq 2$.  A pair of
spin-1 bosons,
however, can make $\hat{\cal V}_{ij}=0$ with $\ell=1$ by making its spin
part antisymmetric, i.e., forming a spin-1 state $\overline{\int}
(u_{1}-u_{2})\vec{A}^{\dagger}_{1}\times
\vec{A}^{\dagger}_{2}|0\rangle$.  This shows that  angular momentum
states forbidden to scalar bosons will become accessible in the
presence of internal degrees of freedom.  Relative angular
momentum, however, costs kinetic energy. One will then have to
balance potential and kinetic energy.

Among all the ground states, the state with {\rm minimum} angular
momentum $L^{\ast}$ that `quenches' interaction (i.e. making $\hat{\cal
V}=0$) plays a special role. It will be referred to as the `minimal
quenching' state $|\Phi^{\ast}\rangle$. The vanishing of $\hat{\cal V}$
can only occur if no two bosons can occupy the same point in space,
meaning that
$B$ in eq.(\ref{Phi}) contains a Laughlin factor $W[z]\equiv \Pi_{N\geq
i>j \geq 1} (u_{i}-u_{j})$.  For spinless particles,
Bose statistics demand that $W$
appears twice; resulting in a minimal quenching state
$B = W^2$, with $L^{\ast}= N(N-1)$.  For spin-1 bosons,
the condition for quenching means
\begin{equation}
B([u], [\mu]) = W[u]\,  F([u], [\mu]),
\label{B} \end{equation}
where $F([u], [\mu])$ obeys Fermi statistics.
Thus $L^{\ast}=N(N-1)/2+P$, where $P$ is the lowest angular momentum
possible for the fermionic wavefunction $F$.

Since there are only three distinct spin states for spin-1 bosons,
the only clusters whose quenching
`fermion' component $F$ has $L=0$  are
$N=2,3$. For $N=2$,  we have
$|\Phi^{\ast}\rangle^{(2)}=\overline{\int}\, u_{12}\,
\vec{A}^{\dagger}_{1}\times\vec{A}^{\dagger}_{2}|0\rangle$,
where $u_{ij}\equiv u_i - u_j$. It has
$(L^{\ast}=1, S=1)$. For
$N=3$, we have  $|\Phi^{\ast}\rangle^{(3)}
 = \overline{\int} u_{12} u_{23} u_{31} \vec{A}^{\dagger}_{1}\times
\vec{A}^{\dagger}_{2}\cdot
\vec{A}^{\dagger}_{3} |0\rangle$, with $(L^{\ast}=3, S=0)$.
For $N>3$, the quenching fermionic function
$F$ must have non-zero angular momentum,  for otherwise there will be
two `fermions' occupying the same state.
One must therefore
populate particles in different
spin and orbital states. This procedure gives rise to a
{\em unique} minimal quenching structure, whose general form will be
clear after we enumerate a few cases. Writing
$|\Phi^{\ast}\rangle^{(N)} = \overline{\int} W  D^{(N)} |0\rangle$, we
have
\begin{eqnarray}
&D^{(4)} =& [\vec{A}^{\dagger}_{1}
\cdot \vec{A}^{\dagger}_{2} \times \vec{A}^{\dagger}_{3}]
 ( u_4 \vec{A}^{\dagger}_{4}), \nonumber \\
&D^{(5)} =& [\vec{A}^{\dagger}_{1}
\cdot  \vec{A}^{\dagger}_{2} \times \vec{A}^{\dagger}_{3}]
 ( u_4 \vec{A}^{\dagger}_{4} \times u_5 \vec{A}^{\dagger}_{5} ) ,
 \\
&D^{(6)} = & [\vec{A}^{\dagger}_{1}
\cdot \vec{A}^{\dagger}_{2} \times \vec{A}^{\dagger}_{3}]
( u_4 \vec{A}^{\dagger}_{4} \times  u_5 \vec{A}^{\dagger}_{5}
\cdot u_6 \vec{A}^{\dagger}_{6}),       \nonumber \\
&D^{(7)} =&  [\vec{A}^{\dagger}_{1}
\cdot \vec{A}^{\dagger}_{2} \times  \vec{A}^{\dagger}_{3}]
[ u_4 \vec{A}^{\dagger}_{4}\times u_5 \vec{A}^{\dagger}_{5}
\cdot u_6 \vec{A}^{\dagger}_{6}]
(u_{7}^{2} \vec{A}^{\dagger}_{7}),\nonumber
\end{eqnarray}
which gives
$(L^{\ast}=7, S=1)$, $(L^{\ast}=12, S=1)$, $(L^{\ast}=18,
S=0)$,  and $(L^{\ast}=26, S=1)$ for $N=4,5,6$ and 7 respectively.
In contrast, the corresponding $L^{\ast}$ for scalar bosons
are  12, 20, 30, 42 , about 100$\%$ higher.  Since $W$ is antisymmetric,
only the antisymmetric part of $D^{(N)}$
will appear in $|\Phi^{\ast}\rangle^{(N)}$.
The antisymmetrization turns the operator product into a determinant,
just as fermion systems\cite{anti}.  This prescription
yields spin $S=0,1,1$ and quenching
angular momentum $L^\ast=2 N^2/3-N+S$
for $N\equiv 0,1,2 ({\rm mod} 3)$.

Note that as particle number increases, more and more singlets made up of
 boson triplets appear in the minimal quenching state.
This is a clear sign that the system becomes a spin liquid
while evolving into a full fledged quantum Hall state.
Note  also that the Laughlin factor $W$ forces all ground states with
$L>L^{\ast}$ to have zero interaction energy.

{\bf III. The route to quenching:}
We now turn to the ground states with $L<L^{\ast}$.
To illustrate the
correlation between $S$ and $L$,
we denote the spin of the ground state
of
$\hat{\cal V}$  with angular momentum $L$ as $S_{L}$, and display their
value
up to  $L^{\ast}$  for a cluster of $N$ bosons  as
$[S_{L}]_{N} \equiv (S_{0},
S_{1},  ..., S_{L^{\ast}})_{N}$. For $c_{2}>0$, we have
$(0,1)_{2}$; $(1,1,1,0)_{3}$,  $(0,1,0,1,$ $ 0,1,1,1)_{4}$,
$(1,1,1,$ $1,1,1,$ $1,1,1,$ $1,0,0,$ $1)_{5}$
; and for
$c_{2}<0$, we have $(2,1)_{2}$,  $(3,2,1,0)_{3}$,
$(4,3,2,1,$ $2,1,1,1)_{4}$,
$(5,4,3$, $3,1,1$, $3,1,1$, $1,2,2$, $1)_{5}$.

To understand these  structures, we note that
since $L<L^{\ast}$, there is not enough angular momentum to produce a
Laughlin factor to prevent the  coincidence of any two bosons. Still,
it is desirable to generate as many boson pairs with {\em relative}
angular momentum  of unit strength as possible.  This consideration
then motivates some simple rules for possible candidates of the
ground state, and for small clusters, they often narrow down to the
exact structure. The rules are:  (i) For given $L$, the ground state of
interaction $\hat{\cal V}$  will contain a maximum number of pairs with
unit {\em relative} angular momentum. (ii) For $c_{2}>0$, (or $c_{2}<0$)
the ground state of $\hat{\cal V}$ will have a  minimum (or maximum)
total spin consistent with rule (i) and Bose statistics. We have
constructed states according to these rules for clusters up to $N=4$
bosons.
The results are given in Table 1, which reproduces the
sequences $[S_{L}]^{(4)}$ listed above.    When compared
with the exact numerical results in the limit
$|c_{2}|/c_{o}<<1$, we found that most of states in Table 1
 coincide with the exact numerical results, or otherwise have
over $94\%$ overlap with them.
 Since the interaction
$\hat{\cal V}$ is invariant under separate spin and orbital
rotation, the correlation between $S$ and $L$ comes entirely from
statistics, just as the  Hund's first and second rule.

In figure 2, we  show the densities of a $N=5$ clusters with
$L=0,3,6,9,12$, and the spin densities for the cluster with
$N=5, L=3, S=1$. The emergence of a
quantum Hall plateau in the density is clear near the
quenching limit. (See also Section {\bf V}).

{\bf IV.  The ground states as a function of $\Omega$:}
Experiments
performed at fixed rotational frequency $\Omega$ are described by the
Hamiltonian
${\cal H}\equiv  H-\Omega L_{z}$. The ground state is given by the minimum
of
${\cal H}_{L} = {\cal V}_{L} +
\hbar(\omega-\Omega)L$, where ${\cal  V}_{L}$ is the ground state energy of
$\hat{\cal V}$ with $L_{z}=L$. If $L$ were a continuous variable, and
${\cal V}_{L}$ a smooth curve, the optimum value $L_{o}$ will satisfy
$\hbar(\omega-\Omega) = -\partial {\cal V}_{L}/\partial L$, and
$\partial^2 {\cal V}_{L}/\partial^2 L >0$.  In addition, we have
$\partial \Omega/\partial L_{o} = (\partial^2 {\cal V}_{L}/\partial^2
L)_{L_{o}}$.  From figure 1,  we see that
aside from  a small number of cusps (in this case at $L=5$ and $7$)
the envelop of ${\cal V}_{L}$ is smooth and concave up, implying that
as $\Omega$ is increased the angular momentum increases in small (but
discrete) steps, reaching $L^*$ at a critical frequency $\Omega^*$.


For $\Omega>\Omega^*$
the minimum of ${\cal H}_{L}$ within
the interval $(0<L<L^{\ast})$ still occurs at $L^{\ast}$. However, over the
interval $L>L^{\ast}$, we have ${\cal V}_{L}=0$, and hence 
${\cal H}_{L} =  \hbar(\omega - \Omega)L$  has a minimum at
$L^{\ast}$.  This establishes Statement {\bf (D)} in the opening, that
for sufficiently rapid rotation the angular saturates at $L^\ast$.

{\bf V. The spin and quantum Hall signature of rotating clusters:}
Because of the large
number ($Q$) of identical but phase incoherent clusters
in the insulating phase, many quantities measured for the entire group of
clusters coincide with the {\em quantum mechanical}
averages of the same quantity within a single cluster, except that it is
 magnified by $Q$.
This is best illustrated by considering
the density of the entire  group of clusters
after the trap is turned off.  Because of tight trapping, the expansion is
mainly driven  by the localization energy of the trap and is therefore
ballistic. The field operator for the cluster (say at
${\bf R}=0$) then evolves as
$\hat{\psi}_{\mu}({\bf r}) = \int\! {\rm d}{\bf r'}\c
G({\bf r} -{\bf r'};t) \hat{\psi}_{\mu}({\bf r'})$, where $G({\bf r};t) =
\prod_{i=1}^{3}\tilde{G}(r_{i};t)$, $\tilde{G}(x;t)=
(M/2\pi i \hbar t)^{1/2}$ ${\rm exp}[iMx^2/2\hbar t]$.
Before  the expansion, the field operator is
$\hat{\psi}_{\mu}({\bf r}) =
\sum_{\ell} w_{\ell}({\bf r}_{\perp}) f_{0}(z) a_{\ell \mu}$, and
the single
particle density matrix of the cluster is
\begin{eqnarray}
\rho_{\mu\nu}({\bf r})&\equiv&  \langle
\hat{\psi}^{\dagger}_{\nu}({\bf r})
\hat{\psi}_{\mu}({\bf r})\rangle =
g_{\mu\nu}({\bf r}_{\perp}) v(z) ,
\label{gw} \\
g_{\mu\nu}({\bf r}_{\perp}) &=& \textstyle
\sum_{\ell}|w_{\ell}({\bf
r}_{\perp}) |^2 \langle a^{\dagger}_{\ell \nu}a_{\ell \mu}\rangle,
\end{eqnarray}
with $v(z) = |f_{0}(z)|^2$, and ${\bf r}_{\perp} = (x,y)$.
In the insulating regime, the density matrix of the entire
cluster collection is $\rho^{\rm all}_{\mu\nu}({\bf r})=
\sum_{\bf R} \rho_{\mu\nu}({\bf r}-{\bf R})$, where the sum is over all
clusters.
After the expansion, we have
$\hat{\psi}_{\mu}({\bf r}, t)$ $=$ 
$\sum_{\ell}$ $\overline{w}_{\ell}({\bf r}_{\perp})$
$ \overline{f}_{0}(z) $ $ a_{\ell \mu} $, 
$\overline{w}_{\ell} ({\bf r}_{\perp})$ $=$ $\int\! $
$ \tilde{G}(x-x';t)$ $ \tilde{G}(y-y';t)$
$w_{\ell}({\bf r}_{\perp}')$ ${\rm d}{\bf r'}_{\perp},$ and
$\overline{f}_{0}(z)$ $=$ $\int\!$ 
$\tilde{G}(z-z';t)$ $ f_{0}(z')$ ${\rm d}{\bf x'}$. 
This implies
$\rho_{\mu\nu}({\bf r}, t)$ $\equiv $$\langle$
$\hat{\psi}^{\dagger}_{\nu}({\bf r} t)$
$\hat{\psi}_{\mu}({\bf r}t)\rangle$ $ =
$ $\overline{g}_{\mu\nu}({\bf r}_{\perp}) $ $\overline{v}(z) $,
where $\overline{g}_{\mu\nu}({\bf r}_{\perp})$ and
$\overline{v}(z)$
are  $g_{\mu\nu}({\bf r}_{\perp})$ and
$v(z)$ in eq.(\ref{gw}) with $w_{\ell}$ and $f_{0}$ replaced by
$\overline{w}_{\ell}$ and $\overline{f}_{0}$.
 Evaluating $\overline{g}_{\mu\nu}$
and $\overline{v}$, we find that $\overline{g}_{\mu\nu}({\bf r}) =
\lambda(t)^{-2}g_{\mu\nu}({\bf r}/\lambda(t))$,
$\overline{v}(z) = \lambda(t)^{-1} v(z/\lambda(t))$,
where $\lambda(t)=\sqrt{1+ (\omega t)^2}$.   We then have
\begin{equation}
\textstyle
\rho^{\rm all}_{\mu\nu}({\bf r},t)
= \lambda(t)^{-3}  \sum_{\bf R} \rho_{\mu\nu}([{\bf r}-{\bf R}]
/\lambda(t) )
\label{scaling} \end{equation}
The diagonal elements of eq.(\ref{scaling}) give the density of
each spin component.
Eq.(\ref{scaling}) shows that the density after expansion is simple a sum of
all the expanded clusters, each of which obeys a scaling
relationship.
For times $t>> 1/\omega $, each cluster has
expanded to a size much larger than the dimension $(\Delta)$ of the origin
cluster ensemble,  eq.(\ref{scaling}) is then well approximated by
$\rho^{\rm all}_{\mu}({\bf r},t) =
Q/(\omega t)^3 \rho_{\mu}\left({{\bf
r}}/{\omega t}\right) $ up to
a correction of order ${\cal O}\left({\Delta}/{a\omega
t}\right )$. The density of each
spin component of the entire expanded ensemble  at
long times  therefore reproduces that of each individual cluster.
(For magnetic field effects, see \cite{last}). 

The identity
$\sum_{\mu}$$\int $$\hat{\psi}^{\dagger}_{\mu}({\bf r})$
$ (r/a)^2 $$\hat{\psi}_{\mu}({\bf r})$
$=\hat{L}_{z} + \hat{N}$ in the LLL implies that if
$L$ is the angular momentum of a cluster before
expansion, then the mean square radius at time $t>>1/\omega$
is $\overline{r^2}=\int\! r^{2}_{\perp}\, \rho^{\rm all}({\bf r},t)
=Qa^2(\omega t)^2 (L+N)$, and
$\overline{z^2}=\int\! z^{2}_{\perp}\, \rho^{\rm all}({\bf r},t)
=Q a^2(\omega t)^2 N/2$. One can therefore extract
the angular momentum of the system from the shape of the entire
expanded clusters.
Moreover, the difference of mean squared radius between successive $L$
states as $\Omega$ increases is
$Q a_{\perp}^2 (\omega t)^2 $.

We would like to thank Immanuel Bloch for discussions. This work is
supported by NASA Grants NAG8-1441, NAG8-1765, and by NSF Grants
DMR-0109255, DMR-0071630.

{\em Notes added:} Concurrent with the submission of this paper, a
preprint (cond-mat/0203061) by B. Paredes, P. Zoller, and I. Cirac
has appeared. Apart from the general discussion of fermionization,
there are little overlaps between our papers.

\vspace{0.2in}

\begin{tabular}{|l||l||l|l||l|l|l|} \hline
N & L  & S  & $D^{ (N)}_{S}$,  $c_{2}>0$  &
 S  & $D^{(N)}_{S}$, $c_{2}<0$ & ${\varphi}_{L}$ \\ \hline

2 & 0 & 0 & $\Theta_{12}$ &
 2 & $A_{1+} A_{2+} $ & 1 \\ \hline

 & $1^{\ast}$ & 1 & $B_{12}$ &
 1 & $B_{12} $ & $u_{12}$\\ \hline \hline

3 & 0 & 1 & $\Theta_{12} \vec{A}_{1}$ &
 3 & $A_{1+} A_{2+} A_{3+}$ & 1 \\ \hline

 & 1 & 1 & $\vec{B}^{}_{12} \times \vec{A}^{}_{3}$  &
 2 & $B^{}_{12 + }  A^{}_{3+}$  &$ u_{12}$\\ \hline

 & 2 & 1 & $\Theta_{12} \vec{A}_{1}  $  &
 2 &  $\Theta_{12} \vec{A}_{1}  $  &$ u_{12}u_{13}$\\ \hline

 & $3^{\ast}$ & 0 & $\Gamma_{123}$  &
0 &   $\Gamma_{123}$  &$w_{123}$ \\ \hline \hline

4 & 0 & 0 & $\Theta_{12} \Theta_{34} $ &
 4 & $\prod_{i=1}^{4}A_{i+}$ & 1 \\ \hline

 & 1 & 1 & $\vec{B}_{12} \Theta_{34} $ &
 3 & $B_{12 +} A_{3+}A_{4}$ & $u_{12}$ \\ \hline

 & 2 & 0 & $\vec{B}_{12} \cdot \vec{B}_{34} $ &
 2 & $B_{12+} B_{34+} $ & $u_{12}u_{34}$ \\ \hline

 & 3 & 1 & $\Gamma_{123} \vec{A}_{4} $ &
 1 & $\Gamma_{123} \vec{A}_{4} $ & $w_{123}$ \\ \hline

  & 4 & 0 & $\vec{B}_{12}\cdot \vec{B}_{34} $ &
 2 &  $B_{12+}B_{34+} $ & $\begin{array}{c} u_{12}u_{14}u_{23} \\
            \times (u_{24}, u_{34})\end{array}$ \\  \hline

  & 5 & 1 & $\Gamma_{123} \vec{A}_{4} $ &
 1 &  $\Gamma_{123}\vec{A}_{4}$ & $w_{123}u_{24}u_{34}$ \\ \hline

  & 6 & 1 & $\Theta_{12}\vec{B}_{34}$ &
 3 &   $A_{1+}A_{2+} B_{34+} $

& $\begin{array}{c} u_{12}^2 u_{13}u_{14} \\
            \times u_{24} u_{34}\end{array}$ \\ \hline

  & $7^{\ast}$ & 1 & $\Gamma_{123}(u_{4}\vec{A}_{4}) $ &
 1 &   $\Gamma_{123}(u_{4}\vec{A}_{4}) $ & $w_{1234}$  \\ \hline
\end{tabular}

\vspace{0.2in}

\noindent Table 1: The ground states constructed from the rules in Section {\bf
III}.
The ground state is written as
$|\Phi\rangle=
\overline{\int}
\varphi_{L} D^{(N)\dagger}_{S}|0\rangle$, using
  $\Theta_{12}\equiv$$ \vec{A}_{1}\cdot \vec{A}_{2}$,
$\Gamma_{123}\equiv$$\vec{A}_{1}$$\times$$\vec{A}_{2}\cdot \vec{A}_{3}$,
$\vec{B}_{12}\equiv$$ \vec{A}_{1} \times \vec{A}_{2}$,
$A_{+}\equiv A_{x} + iA_{y}$,
$w_{_{12..N}}$$\equiv$$\prod_{N\geq i>j\geq 1}$$(u_{i}-u_{j})$,
$u_{ij}\equiv u_{i}-u_{j}$.  All states listed are exact eigenstates
except those with $L=4,5,6$. For $L=4$,  the last factor of
$\varphi$ can be either
$u_{24}$ or $u_{34}$.  The exact ground state is a combination of both.
The state of $L=5$ and 6 has 99.6$\%$ and 94$\%$
overlap with the exact ground state.

\vspace{0.2in}

\psfrag{E}[br]{\small ${\cal V}_L$}
\psfrag{L}[tl]{\small $L$}
\psfrag{VL}[bl]{\small $\rho/\rho_0$}
\psfrag{LL}[cl]{\small $r/a$}
\hspace{0.1in}\begin{tabular}{ccc}
\parbox[tl]{0.5\columnwidth}{
1)
\parbox[t]{0.4\columnwidth}{\verb+ +\\
\includegraphics[width=0.4\columnwidth]{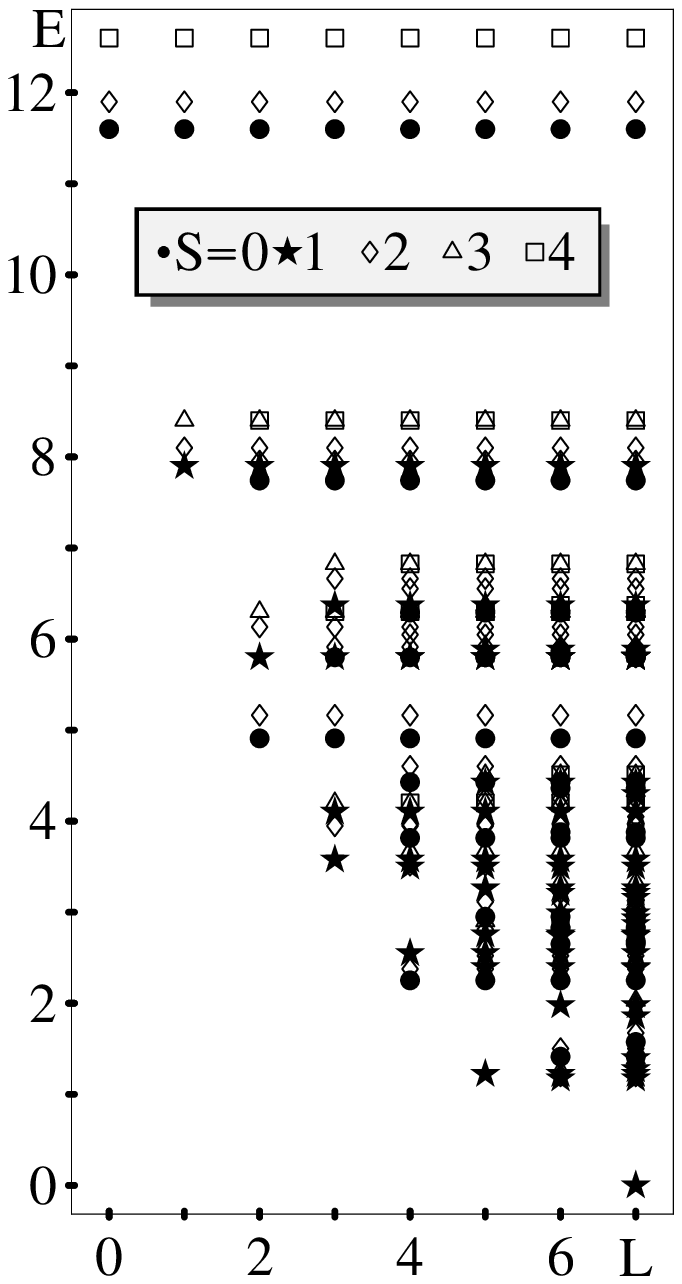}}
\\
{ }}
&\hspace{0.1in}&
\begin{tabular}{c}
2a)\parbox[t]{0.35\columnwidth}
{{\verb+ + }\\\includegraphics[width=0.35\columnwidth]{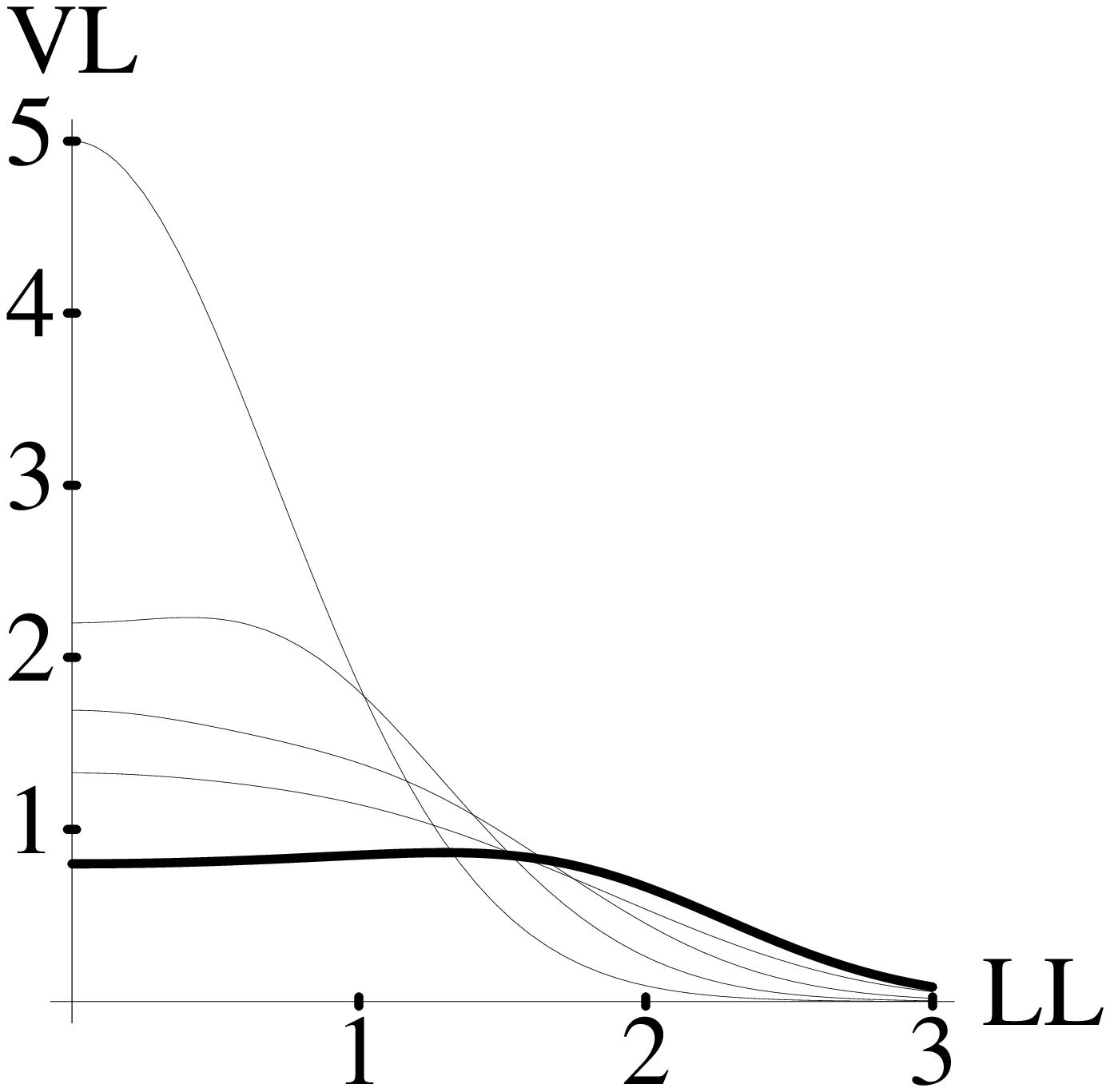}}
\\
2b)
\parbox[t]{0.35\columnwidth}{\verb+ +\\
\includegraphics[width=0.35\columnwidth]{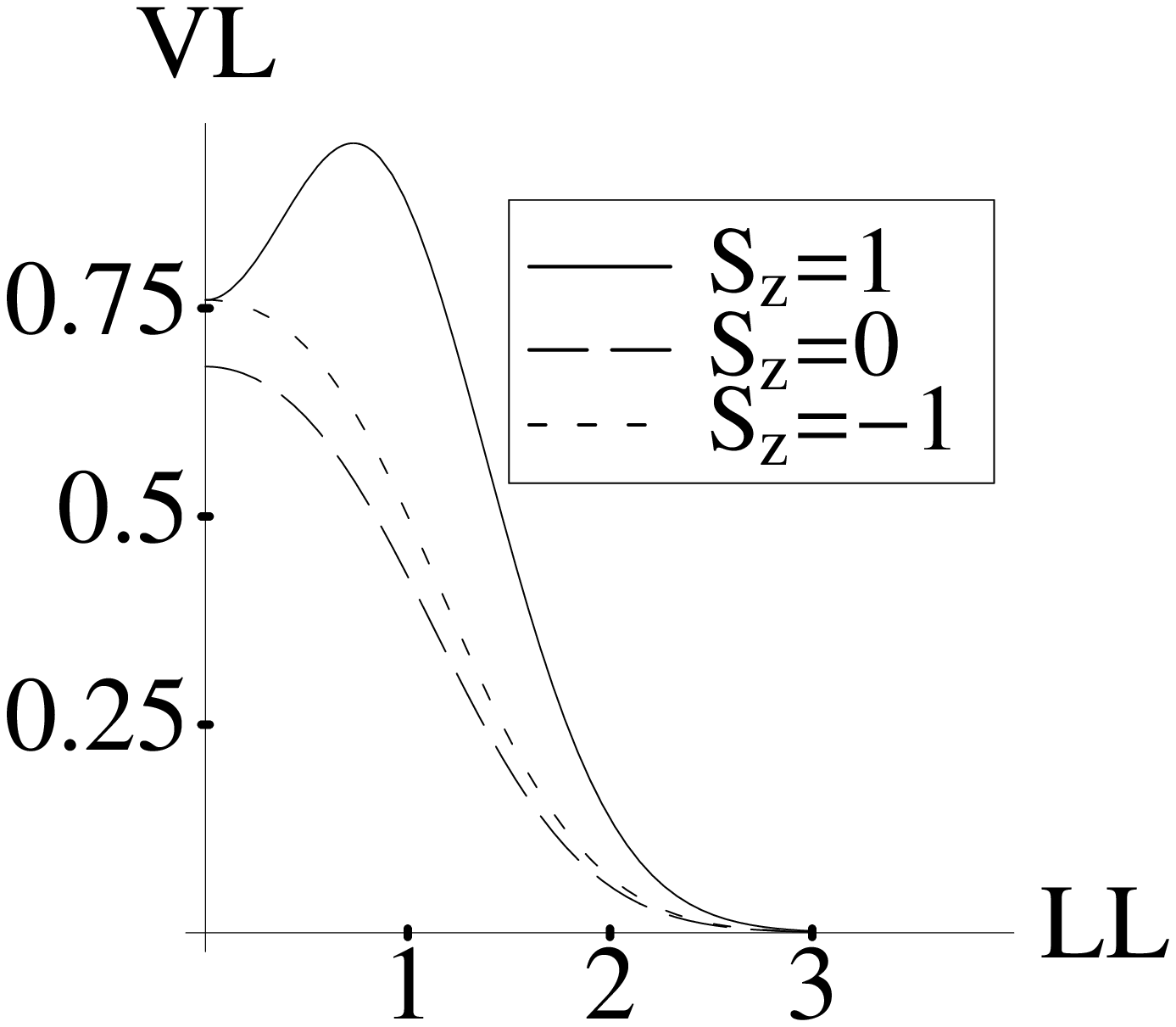}}
\end{tabular}
\end{tabular}

\noindent Figure 1: ${\cal V}_{L}$ for an $N=4$ cluster
with $c_{2}>0$.

\noindent Figure 2a: The column densities of a $N=5$ cluster with
with $L=0,3,6,9,12$.  The central density of these clusters decreases
with increasing $L$.  The $L=12$ data is in bold.
Figure 2b: The spin densities of the
$L=3$ cluster in fig.2a.; ($S=1$).  $\rho_0=1/\pi a^2$.


\begin{thebibliography}{99}
\bibitem{stable} We shall consider $\Omega<\omega$ for otherwise the system
will not be stable.
\bibitem{anti} For example, the antisymmetrization of $D^{(7)}$,
denoted as $\tilde{D}^{(7)}$, is
$\tilde{D}^{(7)}|0\rangle$,
$\tilde{D}^{(7)}=$$\sum_{P} (-1)^{P}$$A^{x \dagger}_{P1}$
$A^{y \dagger}_{P2}$$A^{z\dagger}_{P3}$
$ (u_{_{P4}} A^{x\dagger}_{P4})$
$(u_{_{P5}} A^{y\dagger}_{P5})$
$(u_{_{P6}} A^{z\dagger}_{P6})$
$(u_{_{P7}}^2 A^{x\dagger}_{P7})$,
where $P$ is permutation of the particles. The generalization of this
construction to arbitrary cluster is obvious.
\end{thebibliography}
\end{document}